\documentclass[aps,prc,twocolumn,showpacs,amsmath,amssymb,nofootinbib,superscriptaddress]{revtex4-1}
\usepackage{graphicx} % postscript figures
\usepackage{dcolumn}

\begin{document}
\title{High-spin structures of $^{136}_{~55}$Cs$^{}_{81}$}
\author{A.~Astier}
\author{M.-G. Porquet}
\affiliation{CSNSM, IN2P3-CNRS and Universit\'e Paris-Sud, B\^at 104-108,
F-91405 Orsay, France}
\author{G.~Duch\^ene}
\affiliation{Universit\'e de Strasbourg, IPHC, 23 rue du Loess, F-67037 Strasbourg, France}
\affiliation{CNRS, UMR7178, F-67037 Strasbourg, France}
\author{F.~Azaiez}
\altaffiliation{Present address: IPNO,  IN2P3-CNRS and Universit\'e Paris-Sud, 
F-91406 Orsay, France}
\affiliation{Universit\'e de Strasbourg, IPHC, 23 rue du Loess, F-67037 Strasbourg, France}
\affiliation{CNRS, UMR7178, F-67037 Strasbourg, France}
\author{D.~Curien}
\affiliation{Universit\'e de Strasbourg, IPHC, 23 rue du Loess, F-67037 Strasbourg, France}
\affiliation{CNRS, UMR7178, F-67037 Strasbourg, France}
\author{I.~Deloncle}
\affiliation{CSNSM, IN2P3-CNRS and Universit\'e Paris-Sud, B\^at 104-108,
F-91405 Orsay, France}
\author{O.~Dorvaux}
\author{B.J.P.~Gall}
\affiliation{Universit\'e de Strasbourg, IPHC, 23 rue du Loess, F-67037 Strasbourg, France}
\affiliation{CNRS, UMR7178, F-67037 Strasbourg, France}
\author{M.~Houry}
\altaffiliation{Present address: CEA/DSM/D\'epartement de recherches sur la Fusion
Contr\^ol\'ee, F-13108 Saint-Paul lez Durance, France}
\author{R.~Lucas}
\affiliation{CEA, Centre de Saclay, 
IRFU/Service de Physique Nucl\'eaire, F-91191 Gif-sur-Yvette Cedex, France}
\author{P.C. Srivastava}
\affiliation{Department of Physics, Indian Institute of Technology, Roorkee 247667, India}
\author{N.~Redon}
\affiliation{IPNL, IN2P3-CNRS and Universit\'e Claude Bernard, F-69622 Villeurbanne Cedex, France} 
\author{M.~Rousseau}
\affiliation{Universit\'e de Strasbourg, IPHC, 23 rue du Loess, F-67037 Strasbourg, France}
\affiliation{CNRS, UMR7178, F-67037 Strasbourg, France}
\author{O.~St\'ezowski}
\affiliation{IPNL, IN2P3-CNRS and Universit\'e Claude Bernard, F-69622 Villeurbanne Cedex, France}
\author{Ch.~Theisen}
\affiliation{CEA, Centre de Saclay, 
IRFU/Service de Physique Nucl\'eaire, F-91191 Gif-sur-Yvette Cedex, France}

\date{Received: date / Revised version: date}
\date{\hfill \today}

%%%%%%%%%%%%%%%%%%%%%%%%%%%%%%%%%%%%%%%%%
\begin{abstract}
Odd-odd $^{136}$Cs nuclei have been produced in the $^{18}$O + $^{208}$Pb 
and $^{12}$C + $^{238}$U fusion-fission reactions and their 
$\gamma$ rays studied with the Euroball array. The high-spin level scheme has 
been built up to $\sim$4.7~MeV excitation energy and spin $I \sim 16\hbar$ 
from the triple $\gamma$-ray coincidence data.
The configurations of the three structures observed above $\sim$~2 MeV 
excitation energy are first discussed by analogy with the proton excitations identified in
the semi-magic $^{137}$Cs$_{82}$ nucleus, which involve the three high-$j$ orbits lying 
above the $Z=50$ gap, $\pi g_{7/2}$, $\pi d_{5/2}$ and $\pi h_{11/2}$. This is confirmed 
by the results of shell-model calculations performed in this work.
 
\end{abstract} 


\maketitle

%%%%%%%%%%%%%%%%%%%%%%%%%%%%%%%%%%%%%%%%%%%%%%%%%%%%%%%%%%%
\section{Introduction}
Being less bound than its two even-even neighbors, $^{136}$Xe and $^{136}$Ba, 
the odd-odd $^{136}_{55}$Cs$^{}_{81}$ nucleus cannot be populated by $\beta$ decay, at variance 
to most of the odd-odd nuclei in the nuclide chart. Thus, the $^{136}$Cs nuclei 
were produced for the first time by spallation of La and by fission of U, 
induced by 600 MeV proton beams~\cite{ra75,th81}. This led to the identification 
of two states, the ground state with $T_{1/2}$=13.16~d decaying to excited states
of $^{136}$Ba and an isomeric state with $T_{1/2}$=19~s. The latter 
was measured from the Cs x-rays, meaning that the isomeric state
decays to the ground state via one converted transition at least. Moreover 
the spin values of these two states were obtained from laser spectroscopy, 
$I=5 \hbar$ for the ground state and $I=8\hbar$ for the
isomeric state.

Very recently, the excitation energy of the isomeric state was determined thanks to 
the measurement of its decay to the ground state by means of a 518-keV $E3$ 
transition~\cite{wi11}. 
Moreover a very weak branch  (a 413~keV $M4$ transition) populating an intermediate 
state at 105~keV has been also measured. The ground state and the first 
excited state were then assigned to belong to the multiplet of states with the 
$\pi g_{7/2} \nu d_{3/2}$ configuration, while the isomeric level would be the 
8$^-$ state from the $\pi g_{7/2} \nu h_{11/2}$ configuration~\cite{wi11}.

The high-spin structures built above the $I=8\hbar$ isomeric state remain unknown,
while such structures with only one neutron hole in the $\nu h_{11/2}$ orbit, 
i.e., 
the unique high-$j$ orbit lying below the $N=82$ gap, are expected to be simple 
enough to provide important tests for shell-model calculations. A first candidate
for this purpose is $^{134}$I$_{81}$, unfortunately such a test is 
very limited as a unique structure comprising five excited states was measured 
and discussed in this odd-odd nucleus~\cite{li09,co09}.  

In the present paper, we report on the first identification of the high-spin 
states of $^{136}$Cs, produced as fission fragment in two fusion reactions, 
$^{18}$O + $^{208}$Pb and $^{12}$C + $^{238}$U,  and studied with the Euroball 
array. The level scheme has been built up to $\sim$4.7~MeV excitation energy
and up to spin values around $16 \hbar$. The yrast states are first discussed in 
comparison with the proton excitations already identified in 
$^{137}$Cs$_{82}$~\cite{as12b}. Then the predictions from shell-model calculations using 
the SN100PN  effective interaction~\cite{br05} are presented, showing a good 
agreement with experimental results.

%%%%%%%%%%%%%%%%%%%%%%%%%%%%%%%%%%%%%%%%%%%%%%%%%%%%%%%  

\section{Experimental details}
\subsection{Reaction, $\gamma$-ray detection and analysis\label{exp}}
The $^{136}$Cs nuclei were obtained as fission fragments in 
two experiments. First, the $^{12}$C + $^{238}$U reaction was studied at 90 MeV incident 
energy, with a beam provided by the Legnaro XTU Tandem accelerator. Second, the 
$^{18}$O + $^{208}$Pb reaction was studied with a 85 MeV incident 
energy beam provided by the Vivitron accelerator of IReS (Strasbourg). 
The gamma rays were detected with the Euroball array~\cite{si97}. 
The spectrometer contained 15 
cluster germanium detectors placed in the backward hemisphere with 
respect to the beam, 26 clover germanium detectors located 
around 90$^\circ$ and 30 tapered single-crystal germanium detectors 
located at forward angles. Each cluster detector consists of seven 
closely packed large-volume Ge crystals~\cite{eb96} and each 
clover detector consists of four smaller Ge crystals~\cite{du99}.
In order to get rid of the Doppler effect, both experiments 
have been performed with thick targets in order to stop the recoiling nuclei 
(47 mg/cm$^{2}$ for $^{238}$U and 100 mg/cm$^{2}$ for $^{208}$Pb targets, respectively). 

The data of the C+U experiment were recorded in an event-by-event mode with the 
requirement that a minimum of five unsuppressed Ge
detectors fired in prompt coincidence. A set of 1.9$\times 
10^{9}$ three- and higher-fold events was available
for a subsequent analysis. For the O+Pb experiment, a lower trigger condition 
(three unsuppressed Ge) allowed us to register 4$\times 10^{9}$ events with a 
$\gamma$-fold greater than or equal to 3. The offline analysis consisted 
of both multigated spectra and three-dimensional 'cubes' built 
and analyzed with the Radware package~\cite{ra95}.

More than one hundred nuclei are produced at high spin in 
such fusion-fission experiments, and this gives several thousands 
of $\gamma$ transitions which have to be sorted out. Single-gated
spectra are useless in most of the cases. The selection of one 
particular nucleus needs at least two energy conditions, implying 
that at least two transitions have to be known.
The identification of transitions depopulating high-spin 
levels which are completely unknown is based on the
fact that prompt $\gamma$-rays emitted by complementary 
fragments are detected in coincidence \cite{ho91,po96}.
For the reactions used in this work, we have studied many pairs of 
complementary fragments with  known $\gamma$-ray cascades
to establish the relationships between their number of protons 
and neutrons \cite{po04, as12a}. 
This was taken into account for identifying the $\gamma$-ray 
cascades emitted by the  $^{136}$Cs nucleus, as shown in the 
forthcoming section.

%%%%%%%%%%%%%%%%%%%%%%%%%%%%%%%%%%%%%%%%%%%%%%%%%%%%%%%

\section{Experimental results\label{result}}

\begin{figure}[!h]
\includegraphics*[width=6.5cm]{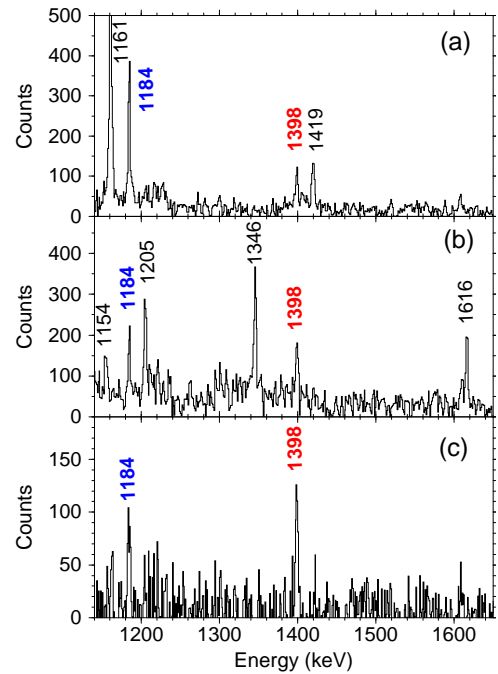}
\caption{(Color online) High-energy part of coincidence spectra gated 
on transitions belonging to the complementary fragments of $^{136}$Cs:  
(a) spectrum double-gated on the 345- and 1227-keV transitions of 
$^{85}$Br~\cite{as06}, built from the $^{18}$O + $^{208}$Pb data set,
(b) spectrum double-gated on the 190- and 331-keV transitions of 
$^{86}$Br~\cite{po09}, built from the $^{18}$O + $^{208}$Pb data set,
(c) spectrum double-gated on the 63- and 84-keV transitions of 
$^{105}$Tc~\cite{lu04}, built from the $^{12}$C + $^{238}$U data set.
The 1184-keV transition (in blue) belongs to $^{137}$Cs~\cite{as12b} and 
the new 1398-keV one (in red) is assigned to $^{136}$Cs. 
}
\label{identification}      
\end{figure}

In order to identify the unknown transitions depopulating high-spin levels 
located above the 8$^-$ isomeric state of the odd-odd $^{136}$Cs isotope, we
have first looked into spectra gated by the first transitions of its main
complementary fragments, $^{85,86}$Br in the O+Pb reaction and  
$^{103-105}$Tc in the C+U reaction. In addition to the $\gamma$ rays known
in $^{85,86}$Br and $^{103-105}$Tc respectively,
new transitions are observed and assigned to $^{136}$Cs. As an example, the spectra of
Fig.~\ref{identification} show the high-energy part of coincidence spectra gated 
on two transitions belonging to $^{85}$Br [spectrum (a)], $^{86}$Br [spectrum (b)], 
$^{105}$Tc [spectrum (c)]. These three spectra exhibit the 1184~keV transition
of $^{137}$Cs~\cite{as12b} and a new transition at 1398~keV which is assigned 
to $^{136}$Cs.
%%%%%%%%%%%%%%%%%%%%%%%%%%%%%%%%%%%%%%%%%%%%%%%%%%%%%%%%%%
%%%%%%%%%%%%%%%%%%%%%%%%%%%%%%%%%%%%%%%%%%%%%%%%%%%%%%%%%%

\begin{figure}[!h]
\includegraphics[width=8.6cm]{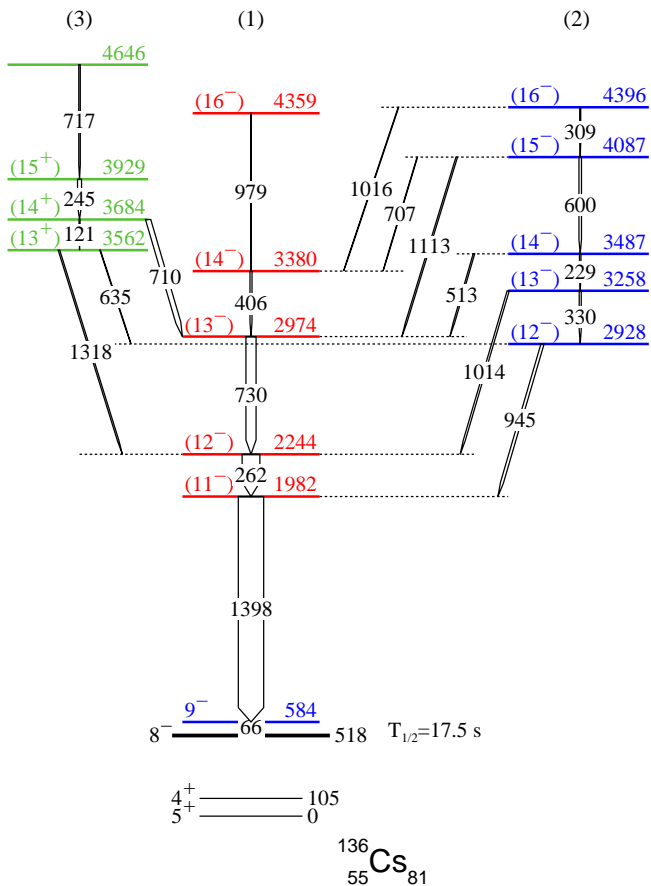}
\caption[]{(Color online) Level scheme of $^{136}$Cs established in this work. 
The excitation energies of the long-lived 8$^-$ state and the 4$^+$ state are 
from Ref.~\cite{wi11}. The width of the arrows is proportional to the 
$\gamma$-ray intensity. The colored states are new. The color code is the same as
that of Fig.~\ref{calculs_SM}.
}
\label{schemaCs136}
\end{figure}

Secondly, we have analyzed spectra in double coincidence with the 1398~keV transition
and one transition of either $^{85,86}$Br in the O+Pb reaction or 
$^{103-105}$Tc in the C+U reaction. This allowed us to identify
unambiguously two other transitions emitted by  the high-spin states of 
$^{136}$Cs, at 262 and 730~keV, which are also weakly observed in 
spectra gated by two transitions belonging to the complementary fragments. Then
the coincidence relationships of the 262-, 730-, and 1398-keV transitions
have been carefully analyzed in order to identify the other transitions
having lower intensity. Finally all the coincidence relationships have been 
looked for. The obtained level scheme is shown in Fig.~\ref{schemaCs136}.
%%%%%%%%%%%%%%%%%%%%%%%%%%%%%%%%%%%%%%%%%%%%%%%%%%%%%%%%%%
%%%%%%%%%%%%%%%%%%%%%%%%%%%%%%%%%%%%%%%%%%%%%%%%%%%%%%%%%%%%%%%%%%%%%%%%%

\begin{figure}[!h]
\includegraphics*[width=7.5cm]{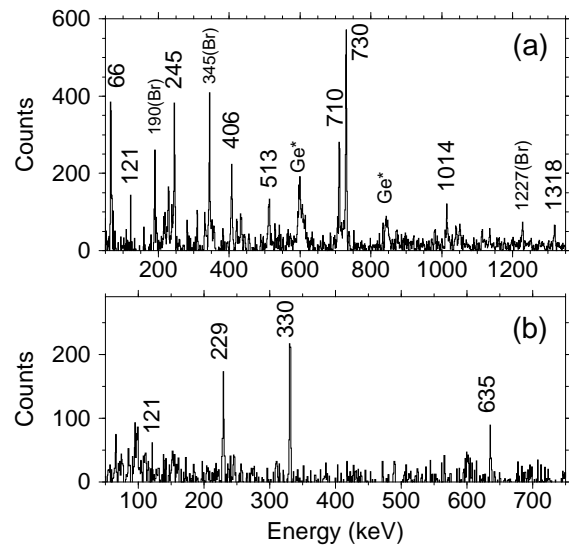}
\caption{Coincidence spectra gated on transitions belonging
to $^{136}$Cs:  
(a) spectrum double-gated on the 1398- and 262-keV transitions, built 
from the $^{18}$O + $^{208}$Pb data set [transitions emitted by $^{85,86}$Br, the
complementary fragments of $^{136}$Cs, are labeled by (Br)],  
(b) spectrum double-gated on the 1398- and 945-keV transitions, built 
from the $^{12}$C + $^{238}$U data set.
}
\label{spectres_yrast}      
\end{figure}

In addition to the three transitions mentioned above, a 66~keV line is observed in all 
the double-gated spectra [see for instance the spectrum shown in Fig.~\ref{spectres_yrast}(a)].
These four transitions in cascade have been located just above the 8$^-$ isomeric state. 
The order of
the four transitions is obtained thanks to their relative intensities measured in various
spectra, in agreement with the existence of several lines which are  
observed in coincidence with only two or three of the four transitions 
(such as the lines at 945, 1014, and 1318 keV). Nevertheless the order of the
66- and 1398-keV transitions could not be disentangled unambiguously, as the uncertainty on 
$I_\gamma(66)$ obtained in spectra gated by transitions belonging to
the complementary fragments is too large.  We have chosen to put the 66-keV
transition at the bottom of the cascade, as discussed below.  
%%%%%%%%%%%%%%%%%%%%%%%%%%%%%%%%%%%%%%%%%%%%%%%%%%%%%%%%%%
%%%%%%%%%%%%%%%%%%%%%%%%%%%%%%%%%%%%%%%%%%%%%%%%%%%%%%%%%%

Above 3~MeV excitation energy, the level scheme comprises two  cascades, in
parallel to the main one. The structure (2) built on the 2928~keV state is very irregular in
energy. All its states decay towards states belonging to the main structure. 
The 3563-keV level is the bottom of a new structure, structure (3), which 
starts with two 
low-energy transitions. It decays mainly towards the 2244-keV state and weakly 
to the 2928-keV one by means of the 635-keV transition. The latter can be seen 
in the spectrum of Fig.~\ref{spectres_yrast}(b).

We have extracted the internal conversion electron coefficients of the 66- and
121-keV transitions by analyzing the relative intensities of transitions 
in cascade. The intensity imbalance of the  66-keV $\gamma$ ray measured 
in spectra in double coincidence with transitions located above it leads to
$\alpha_{tot}$~(66)~=~2.8(10), which is in good agreement with the theoretical
value for a $M1$ multipolarity (see Table~\ref{ICC}). 
%%%%%%%%%%%%%%%%%%%%%%%%%%%%%%%%%%%%%%%%%%%%%%%%%%%
\begin{table}[!h]
\caption{Values of conversion coefficients for the 66.0 and 121.4 keV
transitions, the theoretical values come from the $BrIcc$ 
database~\cite{BrIcc}.}
\label{ICC}
\begin{ruledtabular}
\begin{tabular}{ccccc}
E(keV)\footnotemark[1] &$\alpha _{tot}$(exp)\footnotemark[1] &$\alpha_{tot}^{th}$(E1)\footnotemark[1]
&$\alpha_{tot}^{th}$(E2)\footnotemark[1] &$\alpha_{tot}^{th}$(M1)\footnotemark[1] \\
\hline
 66.0(5)& 2.8(10)& 0.69(2)    & 8.3(3)     & 3.0(1)	\\
 121.4(5)& 0.45(15)& 0.128(2) & 0.94(2) & 0.52(1)	\\
\end{tabular}
\end{ruledtabular}
\footnotetext[1]{the number in parenthesis is the error in the last digit.}
\end{table}
%%%%%%%%%%%%%%%%%%%%%%%%%%%%%%%%%%%%%%%%%%%%%%%%%%%%%%%%%%%%%%%%%%%
Regarding the 121-keV transition, we have obtained 
$\alpha_{tot}$~(121)~=~0.45(10), which is also close to the value expected for 
a $M1$ multipolarity.

As mentioned in the introduction, the 8$^-$ isomeric level of $^{136}$Cs 
belongs to the configuration $\pi g_{7/2} \nu h_{11/2}$ which gives a 
multiplet of states with spin values ranging from 2$^-$ to 9$^-$. This means 
that the 9$^-$ fully-aligned state lies above the isomeric level and would 
belong to the yrast line. Afterwards the increase of angular momentum has to 
involve the breaking of proton pairs. In the neighboring $^{137}$Cs~\cite{as12b} 
the first sign of the breaking is an $E2$ transition of
1184 keV, which is close to the energy of the $2^+_1 \rightarrow 0^+$  
transition of the two even-even isotones, 1313 keV in $^{136}_{54}$Xe and 
1435 keV in $^{138}_{56}$Ba. All these facts would be in favor of putting the
1398-keV transition above the 66-keV one in the yrast cascade of $^{136}$Cs,  
thus defining the spin and parity values of the 584-keV state,
$I^\pi~=~9^-$ (see Fig.~\ref{schemaCs136}). The reverse order of the two 
transitions will be discussed in Sec.~\ref{discuss_SM}.

The statistics of our $^{136}$Cs data is too low to perform $\gamma - \gamma$
angular correlation analyses. Therefore the spin assignments of all the states, 
given in parentheses in Fig.~\ref{schemaCs136}, are based on the following
assumptions: (i) In the yrast decays, spin values increase with excitation
energy and are similar for the states located within the same energy range, 
(ii) Most of the transitions are dipole. In addition, because of the numerous 
links, the parity of structure (2) is assumed to be the same as the one of 
structure (1), i.e. negative. On the other hand, the parity of structure 
(3) is assumed to be positive.

We have gathered in Table~\ref{gammas_Cs136} the properties of all the
transitions assigned to $^{136}$Cs from this work.
%%%%%%%%%%%%%%%%%%%%%%%%%%%%%%%%%%%%%%%%%%%%%%%%%%%%%%%%%%%%%%%%%%%%%%%%%%%%%%%%%%%%%%%%%%%%%%%%%%%
\begin{table}[!h]
\caption{Properties of the transitions assigned to $^{136}$Cs observed in this work.}
\label{gammas_Cs136}
\begin{ruledtabular}
\begin{tabular}{rrccc}
$E_\gamma$\footnotemark[1](keV)&$I_\gamma$\footnotemark[1]$^,$\footnotemark[2]&$I_i^\pi \rightarrow I_f^\pi$&$E_i$&$E_f$\\
\hline
66.0(5)& -\footnotemark[3]&9$^-$ $\rightarrow$ 8$^-$   &583.9 &517.9\footnotemark[4] \\
121.4(5)& 3.3(8)    &   (14$^+$)  $\rightarrow$  (13$^+$)  &3684.1  &3562.6  \\
229.0(4)& 6.0(15)      & (14$^-$)    $\rightarrow$  (13$^-$)  &3486.8  &3257.8  \\
245.1(3)& 15(4)     &  (15$^+$)   $\rightarrow$  (14$^+$)  &3929.2  &3684.1  \\
261.6(2)& 70(10)     &  (12$^-$)   $\rightarrow$ (11$^-$)   &2243.9  &1982.3  \\
309.4(4)& 4(2)      &  (16$^-$)   $\rightarrow$ (15$^-$)   &4396.1  &4086.7  \\
330.2(3)& 8(2)      &  (13$^-$)   $\rightarrow$  (12$^-$)  &3257.8  &2927.6  \\
513.2(4)& 6(2)      &  (14$^-$)   $\rightarrow$  (13$^-$)  &3486.8  &2973.7  \\
406.4(3)& 8(2)      &  (14$^-$)   $\rightarrow$  (13$^-$)  &3380.1  &2973.7  \\
599.8(5)& 9(3)      &  (15$^-$)   $\rightarrow$  (14$^-$)  &4086.7  &3486.8  \\
635.0(5)& 2.5(12)    &   (13$^+$)  $\rightarrow$  (12$^-$)  &3562.6  &2927.6  \\
706.5(5)& 1.0(5)      &  (15$^-$)   $\rightarrow$  (14$^-$)  &4086.7  &3380.1  \\
710.4(3)& 19(5)     &  (14$^+$)   $\rightarrow$  (13$^-$)  &3684.1  &2973.7  \\
716.7(4)& 5(2)      &           $\rightarrow$  (15$^+$)  &4645.9  &3929.2  \\
729.8(3)& 40(8)     &  (13$^-$)   $\rightarrow$  (12$^-$)  &2973.7  &2243.9  \\
945.3(4)& 12(3)     &  (12$^-$)   $\rightarrow$  (11$^-$)  &2927.6  &1982.3  \\
979.1(5)& 2(1)      &  (16$^-$)   $\rightarrow$  (14$^-$)  &4359.2  &3380.1  \\
1013.7(5)& 8(3)     &  (13$^-$)   $\rightarrow$  (12$^-$)  &3257.8  &2243.9  \\
1015.9(5)& 2(1)     & (16$^-$)    $\rightarrow$  (14$^-$)  &4396.1  &3380.1  \\
1113.2(5)& 5(2)     & (15$^-$)    $\rightarrow$  (13$^-$)  &4086.7  &2973.7  \\
1318.5(5)& 5(2)     &  (13$^+$)   $\rightarrow$  (12$^-$)  &3562.6  &2243.9  \\
1398.4(3)& 100(15)   &  (11$^-$)   $\rightarrow$  9$^-$   &1982.3  &583.9  \\
\end{tabular}
\end{ruledtabular}
\footnotetext[1]{The number in parentheses is the error in the last digit.}
\footnotetext[2]{The relative intensities are normalized to $I_\gamma(1398) = 100$.}
\footnotetext[3]{see text}
\footnotetext[4]{from Ref.~\cite{wi11}}
\end{table}
%%%%%%%%%%%%%%%%%%%%%%%%%%%%%%%%%%%%%%%%%%%%%%%%%%%%%%%%%%%%%%%%%%%

\section{Discussion}\label{discuss}
%%%%%%%%%%%%%%%%%%%%%%%%%%%%%%%%%%%%%%%%%%%%%%%%%%%%%%%%%%

\subsection{General features}\label{discuss_gene}
The various configurations of the yrast states of $^{136}$Cs having $I > 8 \hbar$  
can be easily derived from the proton configurations identified in the 
$^{137}$Cs$_{82}$ core, given that the neutron configuration is unique, 
$(\nu h_{11/2})^{-1}$. Several proton configurations have been recently identified in 
$^{137}$Cs~\cite{as12b}, which can be sorted out as a function of (i) the orbit occupied 
by the odd proton and (ii) the number of broken pairs in the $\pi g_{7/2}$ and 
$\pi d_{5/2}$ orbits. They are listed in the left part of 
Table~\ref{config_Cs136}. It is worth recalling that all the fully-aligned 
states indicated in the third column were identified in $^{137}$Cs$_{82}$~\cite{as12b}. 
Regarding the $(\pi h_{11/2})^1$ 
state, the 1867-keV level is a good candidate as it was populated with $L=5$ 
in the proton-transfer reaction $(^3$He, $d$), but its spin value was chosen 
to be 9/2$^-$ because of its decay towards the 7/2$^+$ ground state~\cite{nndc}. 
Nevertheless a $M2$
$11/2^- \rightarrow 7/2^+_{gs}$ transition cannot be excluded, since such a transition is
observed in the neighboring $^{139}$La$_{82}$ isotone~\cite{as12b}.
%%%%%%%%%%%%%%%%%%%%%%%%%%%%%%%%%%%%%%%%%%%%%%%%%%%%%%%%%%
\begin{table}[!h]
\caption{Various proton configurations having zero, one and two 
broken proton pairs in the $\pi g_{7/2}$ and/or $\pi d_{5/2}$ orbit identified 
in $^{137}_{55}$Cs~\cite{as12b,nndc}. By coupling 
the $(\nu h_{11/2})^{-1}$ configuration to each proton one, one obtains the value of
$I^\pi_{max}$(tot) expected for some excited states of $^{136}$Cs.
}
\label{config_Cs136}
\begin{ruledtabular}
\begin{tabular}{cccc|c}
Proton 		&Proton&$I^\pi_{max}$     &$E$(keV)         &$I^\pi_{max}$(tot)\\  
configuration	&seniority&               &$^{137}$Cs	   & $^{136}$Cs\\		
\hline
&&&&\\
$(\pi g_{7/2})^1(\pi g_{7/2}d_{5/2})^4$ &1&7/2$^+$& 0		&9$^-$\\
					&3&15/2$^+$& 1671	&13$^-$\\
					&5&23/2$^+$& 3464	&17$^-$\\
&&&&\\
$(\pi d_{5/2})^1(\pi g_{7/2}d_{5/2})^4$ &1&5/2$^+$& 455		&8$^-$\\
					&3&17/2$^+$&1893 	&14$^-$\\
					&5&21/2$^+$&2783 	&16$^-$\\
&&&&\\
$(\pi h_{11/2})^1(\pi g_{7/2}d_{5/2})^4$ &1&11/2$^-$&1867\footnotemark[1] &11$^+$\\
					 &3&23/2$^-$&3495 	&17$^+$\\
					 &5&31/2$^-$&5492 	&21$^+$\\
\end{tabular}
\end{ruledtabular}
\footnotetext[1]{see text}
\end{table}
%%%%%%%%%%%%%%%%%%%%%%%%%%%%%%%%%%%%%%%%%%%%%%%%%%%%%%%%%%%%%%%%%%%

In all the configurations expected to be involved in the yrast states of 
$^{136}_{55}$Cs$^{}_{81}$, the 
odd nucleons have opposite characters: The neutron is a hole while the protons are 
particles. Thus, the residual interactions lift the degeneracy of each multiplet
so that the states with medium spin values\footnote{The spin of the lowest member of the
multiplet corresponds to the perpendicular coupling of the two angular momenta, i.e., 
$I_{medium}^2(\pi \nu) \sim I_{max}^2(\pi) +
I^2(\nu)$} are lowered in energy as compared to the 
$I_{max}$(tot) state. As a result, the yrast states of $^{136}$Cs are expected to be 
formed from the fragments of each multiplet which comprise the states having spin values 
between $I_{medium}$ and $I_{max}$. 
Then, the states of structure 1 with $I^\pi = (11^-)$ to (16$^-$) and those of structure 
2 with $I^\pi = (12^-)$ to (16$^-$) (see Fig.~\ref{schemaCs136}) have one of the two configurations,  
$(\pi g_{7/2})^1(\pi g_{7/2}d_{5/2})^4$ and $(\pi d_{5/2})^1(\pi g_{7/2}d_{5/2})^4$, with a
proton seniority of 3 or 5. Regarding the states of structure 3, they likely have the 
$(\pi h_{11/2})^1(\pi g_{7/2}d_{5/2})^4$ configuration with a proton seniority of 3.

\subsection{Results of shell-model calculations}\label{discuss_SM}
To get a deeper insight into the configurations of the yrast states of $^{136}$Cs measured
in this work, we have performed shell-model calculations, using the interaction SN100PN
taken from Brown {\it et al.}~\cite{br05} and the NuShellX@MSU code~\cite{br07}. The calculational details are exactly the same as 
those described in Ref.~\cite{wi11} dealing with the energy and decay of the 8$^-$ 
isomeric state of $^{136}$Cs. 

%%%%%%%%%%%%%%%%%%%%%%%%%%%%%%%%%%%%%%%%%%%%%%%%%%%%%%%%%%
\begin{figure*}[!htp]
\includegraphics*[angle=-90,width=17cm]{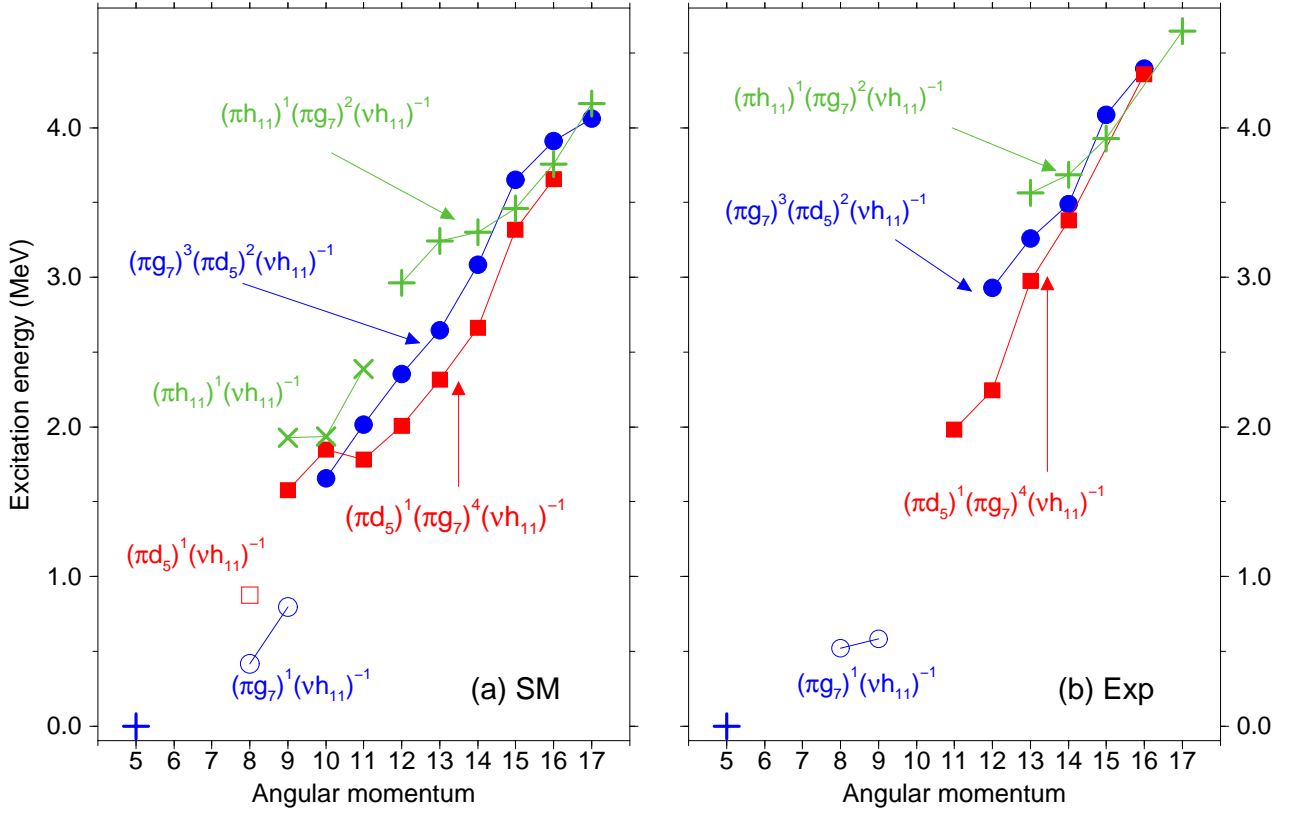}
\caption{(Color online) (a) Excitation energy as a function of angular momentum of the $^{136}$Cs
states having $I \ge 8 \hbar$, predicted by the SM calculations to lie above
the 5$^+$ ground state (see text). The first two or three states for each spin 
value are drawn. The states having the same main configuration are linked by a 
solid line. (b) Experimental states of $^{136}$Cs drawn using the same symbols as in (a).
}
\label{calculs_SM}      
\end{figure*}
%%%%%%%%%%%%%%%%%%%%%%%%%%%%%%%%%%%%%%%%%%%%%%%%%%%%%%%%%%
The left part of Fig.~\ref{calculs_SM} shows the excitation energy above the
5$^+$ ground state (from the $(\pi g_{7/2})^1 (\nu d_{3/2})^{-1}$ configuration) as a 
function of angular momentum of the $^{136}$Cs states with $I \ge 8 \hbar$. Those having 
the same main configuration are drawn with the same symbol and linked by a solid line.

As reported in Ref.~\cite{wi11}, the 8$^-$ isomeric state belongs to the 
$(\pi g_{7/2})^1 (\nu h_{11/2})^{-1}$ configuration. The $I^\pi_{max}=9^-$ state 
lies above it. Then to increase the angular momentum, a proton pair has to be broken. 
At this point, the SM calculation predicts two sets of levels lying between $\sim$1.7~MeV 
and $\sim$4.0~MeV excitation energy. For the yrast states with $I^\pi=11^-$ to 16$^-$ 
[see the red filled squares in Fig.~\ref{calculs_SM}(a)], the odd proton is 
located in the $\pi d_{5/2}$ orbit and one or two $\pi g_{7/2}$ pairs are broken. 
On the other hand, the states having an odd number of protons in the $\pi g_{7/2}$ orbit
have higher excitation energies [see the blue filled circles in Fig.~\ref{calculs_SM}(a)].
Thus the states of structure 1 with $I^\pi = (11^-)$ to (16$^-$) 
(see Fig.~\ref{schemaCs136}) have likely the 
$(\pi d_{5/2})^1(\pi g_{7/2})^4 (\nu h_{11/2})^{-1}$ configuration (noted 1), while the states
of structure 2 with $I^\pi = (12^-)$ to (16$^-$) have the 
$(\pi g_{7/2})^3(\pi d_{5/2})^2 (\nu h_{11/2})^{-1}$ one (noted 2).

It is worth pointing out that, taking into account the fact that the 10$^-$ state of 
configuration 2 is predicted to lie below the 11$^-$ state of configuration 1, one could 
infer that this 11$^-$ yrast state mainly decays towards the 10$^-$ state of configuration 2 
instead of towards the 9$^-_1$ state. Such an assumption would be in agreement with the
experimental results provided that the order of the 66- and 1398-keV transitions is reversed (as
mentioned in Sec.~\ref{result}).
Nevertheless, since the configuration of the 11$^-_1$ state is different from both the 
10$^-_1$ and 9$^-_1$ states, its decay to the lowest state (9$^-_1$), very favored by the 
large transition energy, is more likely, meaning that the 1398-keV transition is located 
above the 66-keV one.  

The fact that the 12$^-$ state of the configuration 2 decays towards the 11$^-$ state 
of the configuration 1 can be explained by the high energy of the transition.  
Thus the predicted  11$^-$ and 10$^-$ states of the 
$(\pi g_{7/2})^3(\pi d_{5/2})^2 (\nu h_{11/2})^{-1}$
configuration could not be observed experimentally in this work. 

The promotion of one proton to the $\pi h_{11/2}$ orbit leads to multiplets of states 
with positive parity. While the predicted $9^+-11^+$ states of the 
$(\pi h_{11/2})^1(\nu h_{11/2})^{-1}$
configuration have a too large excitation energy to be populated in our experiment, 
we can expect to observe the $13^+-17^+$ states due to the breaking of a proton pair, 
i.e., from the $(\pi h_{11/2})^1(\pi g_{7/2})^2 (\nu h_{11/2})^{-1}$ configuration 
with a proton seniority of 3 [see the green plus symbols in Fig.~\ref{calculs_SM}(a)]. 
This is in agreement with the experimental states of structure 3 
(see Fig.~\ref{schemaCs136}).

In conclusion, the use of the SN100PN effective interaction gives a good description of 
the yrast states of the odd-odd $^{136}$Cs. Nevertheless it has to be mentioned that the 
predicted spectrum is slightly more compressed than the experimental one
[compare Figs.~\ref{calculs_SM}(a) and (b)]. As examples, 
the 11$^-_1$ state is calculated(measured) at 1779(1982)~keV, the 13$^+_1$ state
at 3243(3562)~keV. However similar differences are also observed in the 
semi-magic $^{137}$Cs isotope. The 
SM calculations of its excited states have been recently published using the same 
effective interaction~\cite{sr13}. For instance, when selecting levels having the same 
proton-excitation content as the 11$^-_1$ and 13$^+_1$ states of $^{136}$Cs 
(see Table~\ref{config_Cs136} and Fig.~\ref{calculs_SM}), we notice that the 17/2$^+$ 
state is calculated(measured) at 1706(1893)~keV and the 23/2$^-$ state at 3103(3495)~keV. 

\section{Summary}\label{fin}
The odd-odd $^{136}$Cs nuclei have been produced as fission fragments in two fusion 
reactions, $^{12}$C + $^{238}$U and $^{18}$O+$^{208}$Pb, the $\gamma$-rays being
detected  using the Euroball array. The high-spin level scheme  has been
built up to $\sim$~4.7 MeV excitation energy by analyzing triple 
$\gamma$-ray coincidence data. The yrast structures, identified in this work, have 
been firstly discussed in comparison with the general features known in the mass 
region, particularly the proton excitations already identified in $^{137}$Cs$_{82}$, 
which involve the three high-$j$ orbits lying above the $Z=50$ gap, $\pi g_{7/2}$, 
$\pi d_{5/2}$ and $\pi h_{11/2}$. Then shell-model calculations using the SN100PN 
effective interaction have been successfully compared to experimental results.   
They confirm that three high-spin structures observed above $\sim$~2~MeV are due to 
the odd proton lying in one of the three orbits coupled to one or two broken proton 
pairs in the two close orbits ($\pi g_{7/2}$ and $\pi d_{5/2}$), the odd neutron being a
spectator. The SN100PN effective interaction could be further tested using 
the high-spin states of $^{138}_{57}$La. Such a work is in progress.

%%%%%%%%%%%%%%%%%%%%%%%%%%%%%%%%%%%%%%%%%%%%%%%%%%%%%%%%%%  

%%%%%%%%%%%%%%%%%%%%%%%%%%%%%%%%%%%%%%%%%%%%%%%%%%%%%%%%%
\begin{acknowledgments}
The Euroball project was a collaboration among France, the 
United Kingdom, Germany, Italy, Denmark and Sweden. 
The first experiment has been performed under 
U.E. contract (ERB FHGECT 980 110) at Legnaro. 
The second experiment has 
been supported in part by the EU under contract HPRI-CT-1999-00078 (EUROVIV). 
We thank many colleagues for their
active participation in the experiments, Drs. A.~Bogachev, A.~Buta, J.L.~Durell, 
Th.~Ethvignot, F.~Khalfalla, I.~ Piqueras, A.A.~Roach, A.G.~Smith and B.J.~Varley. 
We thank the crews of the Vivitron. 
We are very indebted to M.-A. Saettle
for preparing the Pb target, P. Bednarczyk, J. Devin, J.-M. Gallone, 
P. M\'edina, and D. Vintache for their help during the experiment. 
\end{acknowledgments}
%%%%%%%%%%%%%%%%%%%%%%%%%%%%%%%%%%%%%%%%%%%%%%%%%%%%%%%%%%
%%%%%%%%%%%%%%%%%%%%%%%%%%%%%%%%%%%%%%%%%%%%%%%%%%%%%%%%%%%%%%%%%%%%%%%%%%%%%

\end{document}